\documentclass[aps,onecolumn,10pt, floatfix, superscriptaddress,longbibliography, pra]{revtex4-1} 
\usepackage{graphicx,amsmath}
\usepackage{color}
\usepackage[]{natbib}
\setcitestyle{authoryear,round}
\newcommand{\red}[1]{\textcolor{black}{#1}}

{

\newcommand{\Fr}{\operatorname{Fr}}
\renewcommand{\Re}{\operatorname{Re}}

\usepackage[colorlinks = true,allcolors = {blue}]{hyperref}

\begin{document}

\title{Air cavities at the inner cylinder of  turbulent Taylor-Couette flow}
\author{Ruben A. Verschoof} 
\author{Dennis Bakhuis} 
\author{Pim A. Bullee} 
\author{Sander G. Huisman} 
\affiliation{Physics of Fluids, Max Planck Institute for Complex Fluid Dynamics, MESA+ institute and J. M. Burgers Center for Fluid Dynamics, University of Twente, P.O. Box 217, 7500 AE Enschede, The Netherlands}
\author{Chao Sun}
\email{chaosun@tsingua.edu.cn}
\affiliation{Center for Combustion Energy and Department of Thermal Engineering, Tsinghua University, 100084 Beijing, China}
\affiliation{Physics of Fluids, Max Planck Institute for Complex Fluid Dynamics, MESA+ institute and J. M. Burgers Center for Fluid Dynamics, University of Twente, P.O. Box 217, 7500 AE Enschede, The Netherlands}
\author{Detlef Lohse}
\email{d.lohse@utwente.nl}
\affiliation{Physics of Fluids, Max Planck Institute for Complex Fluid Dynamics, MESA+ institute and J. M. Burgers Center for Fluid Dynamics, University of Twente, P.O. Box 217, 7500 AE Enschede, The Netherlands}
\affiliation{Max Planck Institute for Dynamics and Self-Organization, 37077 G\"{o}ttingen, Germany}

\date{\today}

\begin{abstract} 
Air cavities, i.e.\ air layers developed behind cavitators, are seen as a promising drag reducing method in the maritime industry. 
Here we utilize the Taylor-Couette (TC) geometry, i.e.\ the flow between two concentric, independently rotating cylinders, \red{to study the effect of  air cavities in this closed setup}, which is well-accessible for drag measurements and optical flow visualizations. We show that stable air cavities can be formed, and that the cavity size increases with Reynolds number and void fraction. The streamwise cavity length strongly depends on the axial position due to buoyancy forces acting on the air. Strong secondary flows, which are introduced by a counter-rotating outer cylinder, clearly {\it decrease} the stability of the cavities, as air is captured in the Taylor rolls rather than in the cavity. Surprisingly, we observed that {\it local} air injection is not necessary to sustain the air cavities; as long as air is present in the system it is found to be captured in the cavity. We show that the drag is decreased significantly as compared to the case without air, but with the geometric modifications imposed on the TC system by the cavitators. As the void fraction increases, the drag of the system is decreased. However, the cavitators itself significantly {\it increase} the drag due to their hydrodynamic resistance (pressure drag): In fact, a net drag increase is found when  compared to the standard   smooth-wall TC case.  Therefore, one must first overcome the added drag created by the cavitators before one obtains a net drag reduction.
\end{abstract}

{\let\clearpage\relax\maketitle}

\noindent{\bf Highlights}
\begin{itemize}
\item Study on the effect of air cavities in Taylor-Couette flow
\item Flow visualizations show dependence on Reynolds number, void fraction, and axial position
\item Large difference between net and gross drag reduction
\item \red{Local air injection is not crucial for efficient drag reduction --- as long as sufficient air is available anywhere in the flow.}
\end{itemize}


\clearpage
\subsection{Introduction}
Around 90\% of the world trade is carried by cargo vessels. Therefore, even a minor energy saving in this industry has a major impact on global fuel savings and CO$_2$ emissions. One method to save the overall fuel consumption is to reduce the resistance between the hull and the surrounding water. Next to wave drag and pressure drag (or form drag), viscous skin friction is the major contribution to the total friction, accounting for approximately half of the total resistance \citep{larsson2010}. Wave drag and pressure drag can be optimized by a careful design of the shape of the vessel. Skin friction, however, cannot be optimized similarly, as it is proportional to the wetted area of the hull \citep{Foeth2008}.

One of the most promising techniques in naval engineering to reduce the skin friction is the use of air lubrication. Air lubrication can be applied in the form of bubbly drag reduction (DR) or --- presumably more effective --- in the form of air layer drag reduction \citep{cec10}. Both methods have been studied, mainly experimentally, in great detail over a wide range of Reynolds numbers and bubble diameters, mostly in water channels and flat plate configurations, see e.g.\ the review articles by \citet{cec10} and \citet{murai2014}. The mechanism of bubbly DR is not yet entirely known, but it is clear that large, deformable bubbles are effective in reducing the friction in the boundary layer \citep{gil13, ver16,lu05}. The working principle of DR using air layer is more intuitive: An air layer prevents the water from contacting the hull, thus decreasing the wetted area.  The DR can be over $80\%$, as reported in several studies, as long as the air layer is stable and well-developed \citep{san06, elbing2008, elbing2013, Lay2010}. 

Air layers are formed when sufficient amount of air is injected under a vessel or flat plate \citep{san06, elbing2008}. Although it is a straightforward technique, its drawbacks are the low stability of the air layer and excessive necessary air injection rates, see e.g.\ \citet{Zverkhovskyi2014} for a recent literature overview of air layers and air discharge mechanisms. 

\begin{figure}[htp]
\centering
\includegraphics[scale=1]{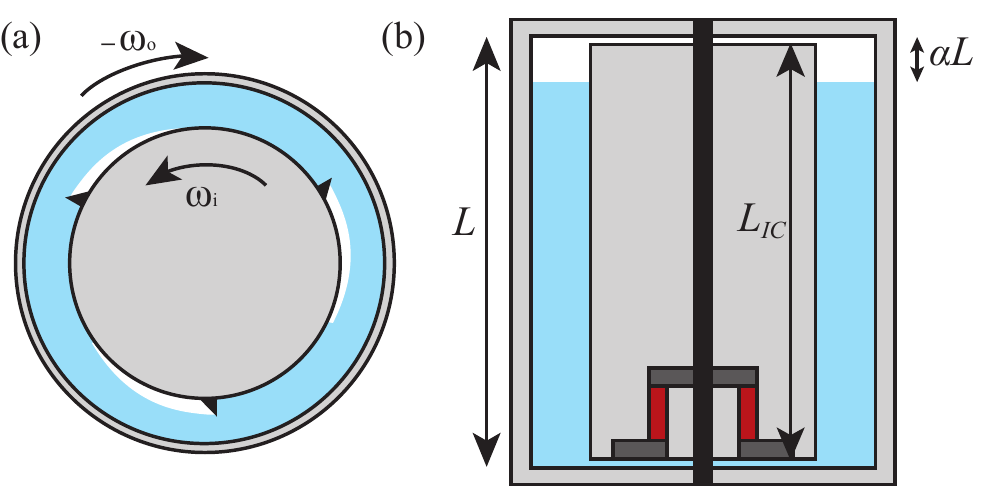}
\caption{{\bf Experimental setup}. {\bf (a)} Top view schematic of the T$^3$C facility (not to scale). Air is captured at the leeward side of the cavitators, as indicated. We attached 2, 3, or 6 cavitators equally distributed around the perimeter of the inner cylinder. The rotation of the cylinders is shown as $\omega_i$ and $-\omega_o.$ (b) Vertical cross-section, showing the position of the torque sensor. \red{The sensor is located in the inner cylinder, so that the torque between the driving shaft and the inner cylinder is measured.} To control the void fraction, we fill the cylinder only partially with water, so that the void fraction $\alpha$ is controlled by measuring the relative height of the water level. \red{Turbulent mixing ensures axial mixing between the two phases, whereas the centrifugal accelerations push the water towards the outer cylinder and, consequently, the air towards the inner cylinder.}}
\label{Chap_Six_fig:setup}
\end{figure}

One way of improvement is by installing a raised edge, commonly referred to as the ``cavitator'', \red{see figures \ref{Chap_Six_fig:setup} and \ref{Chap_Six_fig:cav} for typical cavitator shapes}.  The cavitator creates a region of low pressure at the leeside which stimulates air to attach to the wake of the cavitator, thus decreasing the necessary air injection rate and increasing stability. This air layer developed downstream of the cavitator is called ``air cavity''. Air is usually injected directly at the cavitator, attaches to it, and forms the cavity. Eventually, the air is discharged from the air cavity in the ``closure region'', thus forming a contact line at the wall-water-air interface. In this region, the drag is {\it increased} due to local cavity shedding and re-entrant flows \citep{cec10}. The streamwise air cavity length was found to depend on the gravity wavelength $\lambda$, which is found through the dispersion relation $u_{\infty} = \sqrt{\frac{g \lambda}{2 \pi} \tanh(\frac{2 \pi D}{\lambda})}$, in which $D$ is the water depth, $u_{\infty}$ the free-stream velocity and $g$ the gravitational acceleration  \citep{butuzov1967, matveev2003,matveev2005}. The maximum stable air cavity length then equals half the gravity wavelength $\lambda$, which thus only depends on the water depth and vessel speed . Experiments indeed confirmed that the cavity length is virtually independent on the cavitator height and air injection rate \citep{Zverkhovskyi2014}.  Two different air discharge mechanisms are known in the closure region: i) wave pinch-off and ii) a re-entry jet \citep{cec10}. The wave pinch-off mechanism is related to interfacial waves, that can pinch off patches of air when the air cavity is thin. The re-entry jet is formed by a stagnation point in the wake of the cavity. This jet flows upstream, leading to a periodic break-off of air \citep{Makiharju2013}. Note that the nomenclature of the ``cavitator''  is unrelated to real cavitation, i.e.\ the rapid liquid-to-vapor phase transition.  

When applying air cavities, a significant difference between net and gross DR is present. A compressor to blow air under the hull consumes energy, and locally the drag is increased in the closure region and  by the cavitators and skegs, which are installed to prevent air from discharging sideways. In fact, the challenge is not to increase the drag by the geometric changes, but to find a net drag reduction. Therefore one has to find the optimum between DR and additional energy losses and also expenses. Recent full-scale ship measurements resulted in impressive net power savings of $10\%$ to $20\%$ \citep{Latorre1997,  Hoang2009,Mizokami2010,Amromin2011, Makiharju2012,kumagai2015}. However, despite its clear potential advantages, air lubrication is hitherto not widely used. One of the main issues it that laboratory results are hard to scale up to the conditions of real applications and therefore the performance of ships is difficult to predict \citep{murai2014}. Although efforts are being made to numerically model air cavities, presently only RANS and LES simulations are common, and thus closure models, which are not always reliable, are necessary \citep{Rotte2016}. In this field of research, many questions remain unanswered, and there is a clear need for well-controlled, precise measurements to study the underlying physics.

The goal of this study is to explore the possibilities of studying air cavities in Taylor-Couette (TC) flow. Taylor-Couette flow, i.e.\ the flow between two concentric, independently rotating cylinders, is one of the canonical systems in which fluid flow physics is studied, see the recent reviews by \citet{far14} and \citet{gro16}, and fig.\ \ref{Chap_Six_fig:setup} for a schematic of a TC setup. It has the advantage of being a closed system with an exact balance between driving and energy dissipation, and it is accessible experimentally thanks to its simple geometry. Furthermore, it is a compact system in which highly turbulent flows can be studied. By using a TC apparatus, we have the opportunity to study air cavities in a highly controlled environment. 

In TC flow, the driving of the flow is expressed through two Reynolds numbers, namely $\Re_i=\omega_i r_i (r_o-r_i)/\nu$  for the inner cylinder and $\Re_o=\omega_o r_o (r_o-r_i)/\nu$ for the outer cylinder. Here, $r_i$ and $r_o$ are the radii of the inner and outer cylinder, respectively, $\omega_{i}$ and $\omega_o$ are the angular velocities of the inner and outer cylinder, respectively, and $\nu$ is the kinematic viscosity. The primary response parameter is the torque $\tau$ necessary to rotate the cylinders at a constant driving speed, i.e.\ at constant $\Re_i$ and $\Re_o$. The torque is made dimensionless as \red{$G = {\tau} / (2 \pi L_{IC} \rho \nu^2)$, in which $\rho$ is the fluid density and $L_{IC}$ is the height of the inner cylinder, i.e.\ the height over which the torque is measured}.

Taylor-Couette flow has been used extensively to study bubbly drag reduction \red{ and bubble dynamics}, see e.g.\ \citet{Djeridi2004, climent2007, ber05, ber07, mur05,mur08,sug08b, gil13,cho14,fokoua2015, ver16}. At lower Reynolds numbers, small bubbles do not only beautifully visualize Taylor rolls and other vortical structures in the flow \citep{Ruymbeke2017}, they also decrease the drag by destroying the momentum transport in these vortices \citep{Spandan2016}. At higher Reynolds numbers, in which the drag and shear rates are high, a small percentage of large bubbles has a {\it tremendous} effect on the global drag, e.g.\ at $\Re_i=2\times 10^6$, DR percentages of 40\% were observed for a global gas volume fraction of only $4\%$ \citep{gil13}. These DR percentages reach far beyond the trivial effects of the changed effective density and viscosity.
Efforts are being made to study thin air layers around the inner cylinder in TC flow, which can be achieved by using a superhydrophobic coating \citep{sri15,ros16}, or by heating the inner cylinder, thus creating a Leidenfrost vapour layer \citep{sar16}. Studying air cavities in TC flow has, to the best of our knowledge, never been done before.

The outline of this paper is as follows. First, we describe the experimental method in section \ref{sec:exp}. In section \ref{sec:visu} we show the existence of air cavities by showing and interpreting high-speed recordings. \red{From these visualizations, we extract the cavity length and global coverage in section \ref{sec:cavity_length}.} We continue by presenting the torque measurements and resulting drag reduction in section \ref{sec:torque}, thus quantifying the flow behaviour. We conclude this study in section \ref{sec:conclude}.

\section{Experimental method}
\label{sec:exp}

\begin{figure}[htp]
\centering
\includegraphics[scale=1]{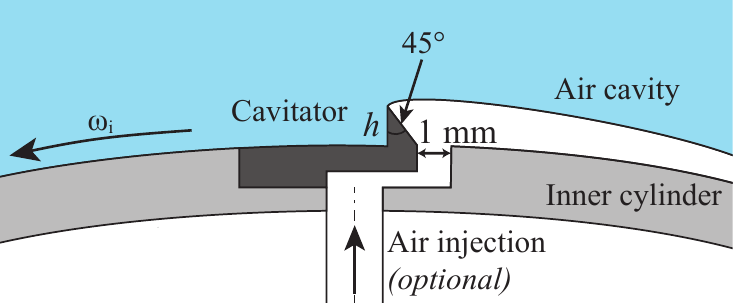}
\caption{Sketch of the cavitator and air injector. The cavitator, here shown in dark gray extends over the entire height of the inner cylinder, see also fig.\ \ref{Chap_Six_fig:sketch_cavity}. The cavitator edge has a sharp, 45$^{\circ}$ corner, with a height of $h =2$  mm. This sharp edge, although introducing pressure drag, is necessary to start a stable air layer. Using a rotary union-slip ring combination, air is led through the shaft, meaning that air can be injected through the inner cylinder while the cylinder is rotating. This way of active air injection is optional.}
\label{Chap_Six_fig:cav}
\end{figure}

The experiments were performed in the Twente Turbulent Taylor-Couette (T$^3$C) facility, see \citet{gil11} for all technical details. The setup has inner and outer cylinder radii of $r_i = 200$ mm and $r_o = 279.4$ mm, respectively, giving a radius ratio of $\eta = r_i/r_o = 0.716$ and a gap width $d=r_o-r_i=79.4$ mm. The maximum rotation frequencies of the inner and outer cylinder with cavitators are $f_i=10$ Hz and $f_o= \pm5$ Hz, respectively, resulting in Reynolds numbers up to $\Re_i = 2\pi f_i r_i d/\nu = 1 \times 10^6$ and $\Re_o = \pm 2 \pi f_o r_o d / \nu = \pm 7 \times 10^5$ with water as the working fluid \red{with a temperature of $T = 20\pm0.5~^{\circ}$C. We continuously measure the temperature, and use the instantaneous temperature-dependent viscosity and density to calculate our dimensionless quantities.} The rotation ratio between outer and inner cylinder is defined as $a=-f_o / f_i$.
When both cylinders rotate, we express the driving as a ``shear Reynolds number'', i.e.\ 
\begin{equation}
\Re_s = \frac{r_i(\omega_i  -  \omega_o) d }{\nu} = \Re_i - \eta \Re_o, 
\end{equation}
in which $\omega_{i,o} = 2 \pi f_{i,o}$. In this study, the rotation rates are limited by vibrations in the system, which are caused by the uneven distribution of air. The \red{outer cylinder has} a height of $L = 932$ mm, giving an aspect ratio of $\Gamma = L/d = 11.7$. \red{The inner cylinder has a height of $L_{IC} = $ 927 mm}. The transparent acrylic outer cylinder allows for flow visualizations. The end plates, which are partly transparent, are fixed to the outer cylinder. The torque $\tau$ is measured with a co-axial torque transducer (Honeywell 2404-1K), placed inside the inner cylinder to avoid measurement errors caused by seal- and bearing friction, see fig.\ \ref{Chap_Six_fig:setup}. All flow visualizations are made with a Photron FASTCAM SA-X high-speed camera with a resolution of $1024 \times 1024$ px. \red{As illumination we used a Briteq BT-Theatre-1EZ LED theatre spotlight. We stress that a uniform light intensity is hard to achieve due to the curved surface of the cylinders. A Zeiss Makro 50 mm lens was used, resulting in a field of view of $23^{\circ}$ in both the horizontal and vertical direction.}
We fix 2, 3, or 6 cavitators to the inner cylinder (IC), as shown in figures \ref{Chap_Six_fig:setup} and \ref{Chap_Six_fig:cav}. Due to a local low pressure, air attaches to it in the wake of the cavitator. The cavitators extend over almost the entire height of the cylinders, and have a height of 2 mm. Shape and size effects of the cavitator have been studied in a water tunnel configuration \citep{Zverkhovskyi2014}, indicating that a sharp cavitator tip is necessary.

\begin{figure}[htp]
\centering
\includegraphics[scale=1.2]{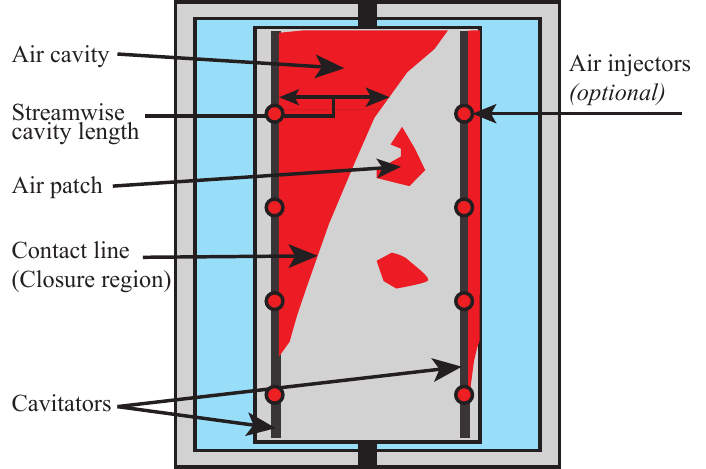}
\caption{A sketch of the front view of the setup with a developed air cavity, which is shown in red to ease readability. The cavitator is shown in dark gray. We indicate the position of all relevant features encountered in this study. The height of the air injectors equals $z/L = {0.09,~0.33,~0.57,~0.81}$.}
\label{Chap_Six_fig:sketch_cavity}
\end{figure}

\red{The amount of air is characterised by the global void fraction $\alpha$.} In the current study, we used 2 procedures of controlling the amount of air in the setup: i) We do fill the apparatus only partially with water, leaving room for a controlled amount of air, \red{which is measured with both cylinders at rest. Already at the lowest Reynolds numbers presented here, }turbulent mixing ensures distribution of air in the bulk flow (see fig.\ \ref{Chap_Six_fig:setup} and e.g.\ \citet{ver16}). \red{This procedure is used, unless mentioned otherwise.} Or ii) We actively inject air from the cavitators at \red{four axially distributed heights}, as indicated in \red{figs.\ \ref{Chap_Six_fig:cav} and \ref{Chap_Six_fig:sketch_cavity}}. An overflow channel allows air to leave the setup and prevents pressure build-up in the system. \red{After each measurement, when both the cylinders and the fluid are at rest again, we remeasure the void fraction, to ensure that the void fraction was constant during the measurement. With this procedure, the uncertainty in void fraction is kept below $\alpha_{err} \leq 0.1\%$. We note that in both procedures, the local effective local void fraction depends on axial and radial position due to buoyancy and centripetal forces, as shown in \citet{gil13}. }

\section{Results}
\begin{figure*}[htp]
\centering
\includegraphics[width=.95\columnwidth]{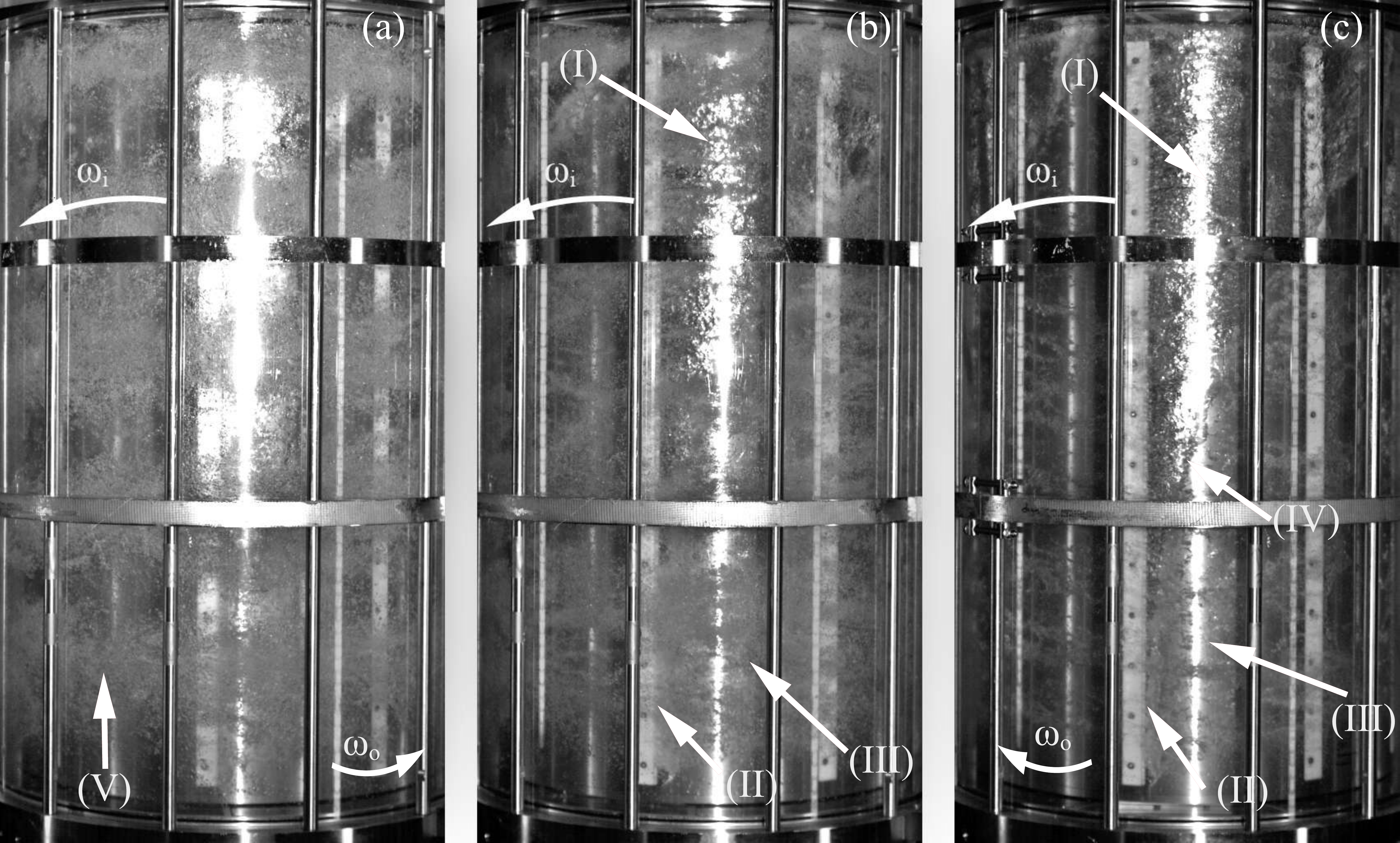}
\caption{\red{ Snapshots of air cavities at $\Re_s = 8\times 10^5$, for 3 different rotation ratios: {\bf (a)} $a=0.14$ (counter-rotation), {\bf (b)} $a=0$ (stationary OC) and {\bf (c)}  $a=-0.2$ (co-rotation). The direction of the cylinder rotation is indicated by the curved arrows, in which $\omega_i$ and $\omega_o$ indicate the direction of the inner and outer cylinder, respectively. The global gas volume fraction is $\alpha=2\%$. The vertical bars and horizontal rings are essential structural parts of the setup. (I) The air cavity. (II) The cavitator. (III) Cylinder not covered with an air cavity. (IV) The contact line of the cylinder-water-air interface. (V) In the counter-rotating case, many bubbles are trapped in these `Taylor vortices'. In fig.\ (b) and especially in fig.\ (c), the flow is radially more stably stratified. Therefore, less bubbles are present and the air cavity is better visible.}}
\label{Chap_Six_fig:visu2}
\end{figure*}
\begin{figure*}[htp]
\centering
\includegraphics[width=0.95\columnwidth]{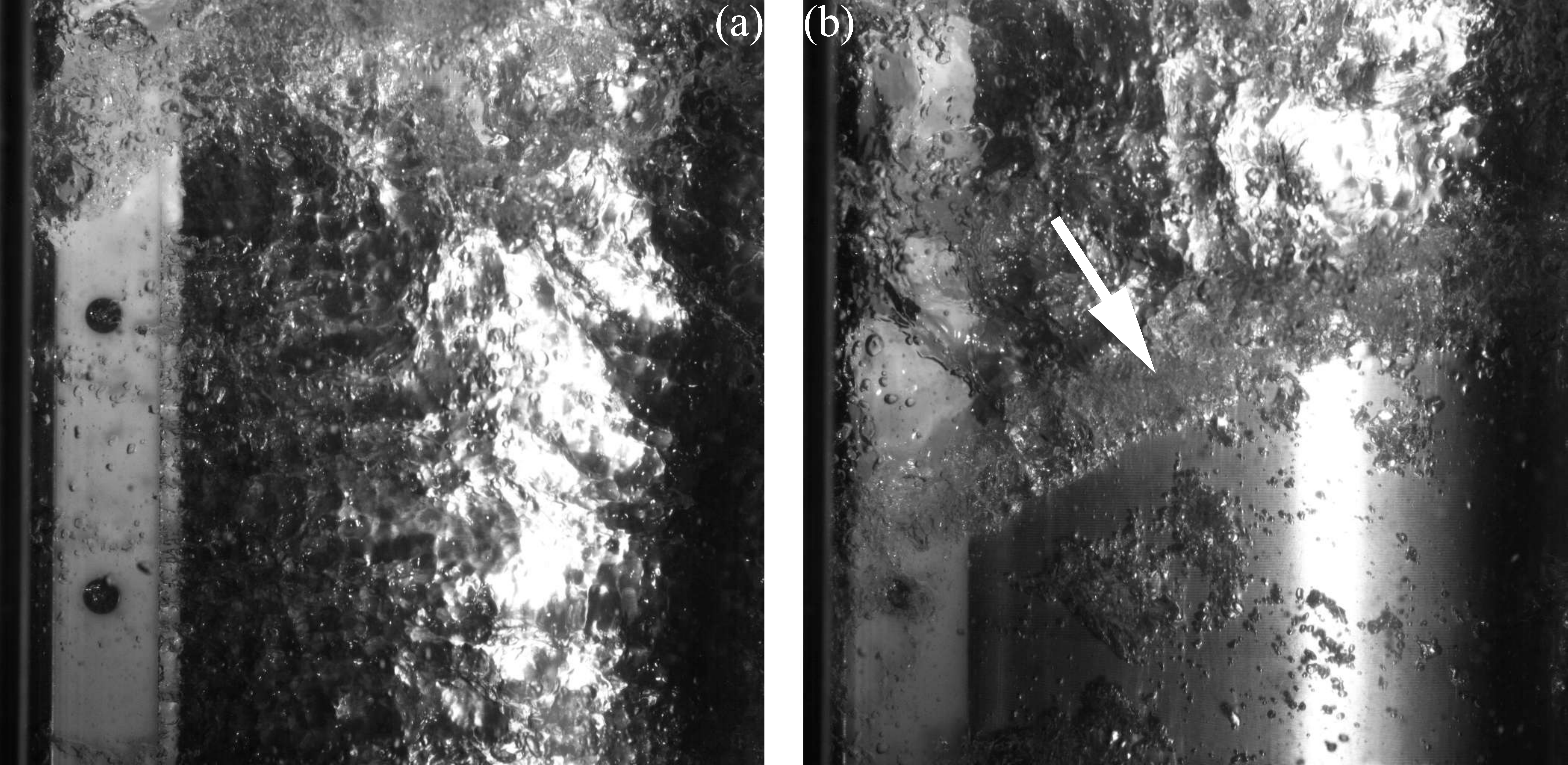}
\caption{Snapshots of air cavities at $\Re_i = 5 \times 10^5$ with a stationary outer cylinder, 3 cavitators, and 2\% of air. We zoomed in on the top of the cylinder. We show 2 photos taken at time = $t_1$ --- when the cavitator is visible, and $t_2$ --- when the closure region is visible. {\bf (a)} Cavitator (vertical white strip in image) with development of the air cavity \red{at $t_1$}. {\bf (b)} The closure region of the air cavity \red{at $t_2$}. The visible white bar here is {\it not} a cavitator, \red{but a blank that is mounted flush with the cylinder surface}. Note the dependence of the cavity length on the height. The white arrow indicates the position of the closure region, which is governed by the re-entry jet mechanism.}
\label{Chap_Six_fig:visu}
\end{figure*}

\subsection{Flow visualizations}
\label{sec:visu}

The goal of these visualizations is two-fold: i) we can qualitatively study the flow dynamics to prove the existence of air cavities in TC flow, and ii), we can extract the streamwise air cavity length and air coverage from these images.  
We first visualize the entire TC setup, as shown in fig.\ \ref{Chap_Six_fig:visu2}. The air cavity is best visible in fig.\ \ref{Chap_Six_fig:visu2}c. Here, we see the cavity over a considerable portion of the cylinder, especially in the top part of the setup. We observe \red{the interfacial capillary waves at the surface of the air cavity, and we see }bubbles or air patches at locations where the cavity is not formed. \red{See the annotations in fig.\ \ref{Chap_Six_fig:sketch_cavity} and \ref{Chap_Six_fig:visu2}, in which all relevant flow features are highlighted.}

In fig.\ \ref{Chap_Six_fig:visu2}a the cylinders counterrotate, which is known to induce turbulent Taylor vortices \citep{gil12,ost14pd}. These vortices are visualized here by the bubbles captured within them. The vortices, which introduce strong secondary flows in the system, prevent the bubbles from sticking to the inner wall. Apparently, the air cavity is largely destabilized and destroyed in the counterrotating regime.
In fig.\ \ref{Chap_Six_fig:visu2}c, the opposite is the case. Corotating cylinders stabilise the flow and suppress secondary flows. We see that air is pushed towards the inner cylinder more effectively than in the counterrotating case. Fig.\ \ref{Chap_Six_fig:visu2}b, in which only the inner cylinder rotates, shows an intermediate behaviour, i.e.\ no pronounced Taylor vortices, but nonetheless larger vortical structures in which bubbles are entrained. 

To capture the local dynamics we zoom in, as shown in fig.\ \ref{Chap_Six_fig:visu}. Here, we only rotate the inner cylinder. We clearly see the interfacial waves at the air-water interface. Also, it is clear that the majority of the air is indeed captured in the air cavity --- only few bubbles are present in the bulk. In the closure region, we see that the dominant breakup process is the ``re-entry jet''. This is clear from the local structure, which is more ``blurry'' than most of the cavity, as indicated by the white arrow in fig.\ \ref{Chap_Six_fig:visu}b.

\subsection{Cavity length and coverage}
\label{sec:cavity_length}
\begin{figure}[htp]
\centering
\includegraphics[scale=1.1]{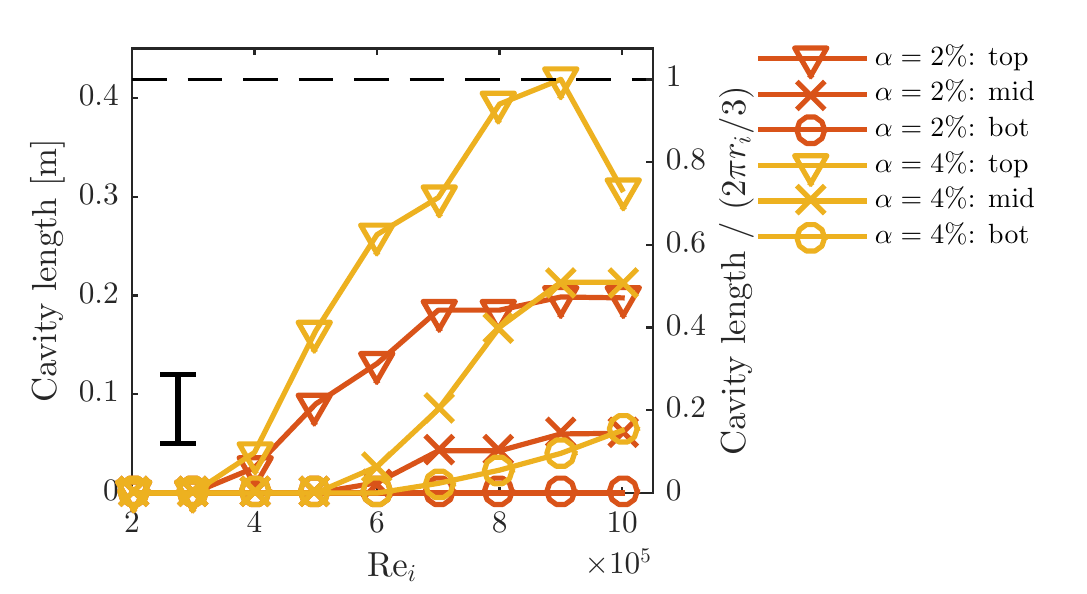}
\caption{Streamwise air cavity length on the inner cylinder as a function of $\Re_i$. The outer cylinder is stationary. We used 3 cavitators. The coverage is extracted by visual means from a series of images similar to those of fig.\ \ref{Chap_Six_fig:visu2}. We show results for three different axial positions, close to the top ($z/L=3/4$), at mid-height ($z/L=1/2$) and close to the bottom ($z/L=1/4$). The estimated error bar is shown \red{in the bottom left corner of the graph. In dashed black, we added the streamwise length between two cavitators $2 \pi r_i/3$, which is the upper limit of the streamwise cavity length. On the right y-axis, we normalized the streamwise cavity length with the distance between two cavitators.} }
\label{Chap_Six_fig:length}
\includegraphics[scale=1.1]{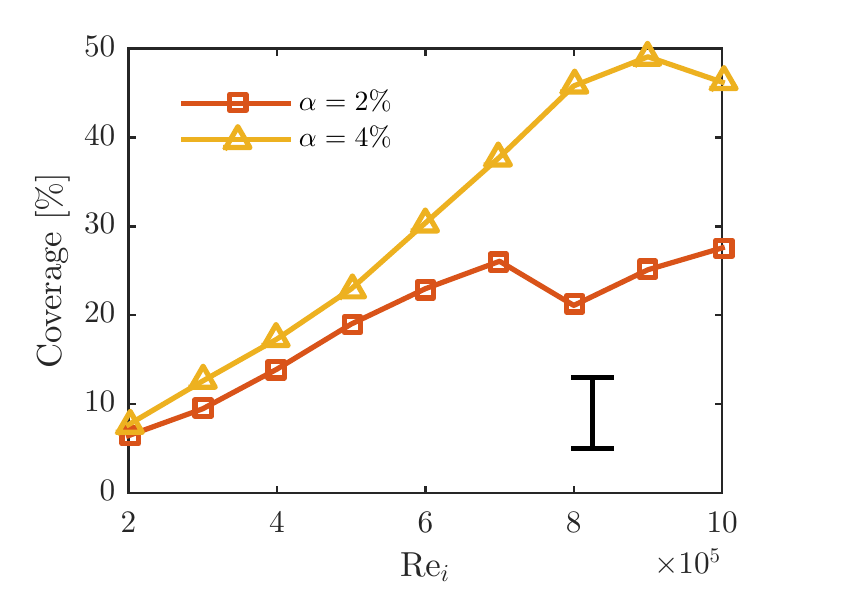}
\caption{Percentage of air cavity coverage on the inner cylinder as a function of $\Re_i$. The outer cylinder is stationary. We used 3 cavitators. The coverage is calculated by integrating the streamwise cavity lengths (see fig.\ \ref{Chap_Six_fig:length}). The estimated error is shown \red{in the bottom right corner of the graph}.}
\label{Chap_Six_fig:coverage}
\end{figure} 
The streamwise cavity length and global coverage are among the crucial parameters of the cavity as they govern the de-wetted area and thus the possible drag reduction. 
 From images as shown in fig.\ \ref{Chap_Six_fig:visu2}, we extract the air cavity length. \red{We first averaged 100 independent instantaneous photos of the flow, to get a time-averaged cavity length rather than the instantaneous value. Then, we manually tracked the edges of the cavity and the cavitator, and binarized the image, in which the area covered by cavity is distinguished from the area which is not. From this binarized image, we can extract the global coverage and the streamwise air cavity length at any axial position. Due to the structural parts blocking the view and light reflections, an automated procedure to extract the position of the air cavity turned out to be unfeasible.} We do this for all cases with 3 cavitators and a stationary outer cylinder.
Here, the cavity length is shown for 3 axial heights in fig.\ \ref{Chap_Six_fig:length}. The axial dependence, which already was made clear from fig.\ \ref{Chap_Six_fig:visu2} and \ref{Chap_Six_fig:visu} is significant. 
Clearly, due to buoyancy forces air has an spatial preference towards the top of the cylinders. Previous bubble DR measurements already showed the axial dependence of the location of bubbles, even when air is continuously injected from the bottom \citep{gil13}. We see here that this axial dependence is present also in the case of air cavities. \red{We note that although the cavity length at the three shown axial heights for the smallest Reynolds numbers is zero, a small cavity already forms closer to the top.}

The centrifugal acceleration at the inner cylinder is given by $a_{centr} = \omega_i^2 r_i$. Consequently, in the limit of $\Re_i \rightarrow \infty$, the gravitational forces are negligible as compared to $a_{centr}$, and the axial dependence of the air cavity length will disappear.
In the hypothetical case of air in a purely laminar TC flow, all air would be pushed towards the inner cylinder due to a radial pressure gradient caused by the centrifugal forces \citep{gil13}. In this turbulent flow, however, strong velocity fluctuations are present, causing the air to distribute itself over the entire gap width, even though the preferential accumulation close to the inner wall remains present \citep{gil13}.
\red{The number of cavitators does not influence the cavity length, as long as the air cavities can be regarded as isolated \citep{Zverkhovskyi2014}. If the cavitators are so closely spaced that cavitators are within range of an upstream cavity, they become submerged in air and a continuous air cavity is formed.}

We calculate the air coverage for both $\alpha=2\%$ and $\alpha=4\%$, to know whether the global gas fraction influences the coverage, see fig.\ \ref{Chap_Six_fig:coverage}. The coverage is calculated by integration of the streamwise cavity lengths over the entire height of the cylinder, which is then divided by the total area between 2 cavitators (i.e.\ $2 \pi r_i L_{IC} / 3$). We see that the coverage for both measured gas fractions are similar up to $\Re_i = 6 \times 10^5$, after which the $\alpha=2\%$-curve saturates at a coverage of $25\%$. The coverage for $\alpha=4\%$ increases up to $\Re_i=8\times10^5$, where it saturates at a coverage of 45\%.

In these 2 saturation coverages regimes, the majority of the air is attached to the cavity, and not dispersed throughout the flow as bubbles. Assuming that all air is attached to the cavity, it is possible to get an estimate of the thickness of the air cavity. The surface area covered of the inner cylinder equals $2\pi r_i L_{IC}\cdot coverage$. We divide the volume of air in the setup, which equals $\alpha V= \alpha \cdot L_{IC} \pi(r_o^2-r_i^2)$, by the coverage surface area, to get a nominal value for the thickness $h_{cavity}$. For both $\alpha=2\%$ and $\alpha=4\%$ we find that $h_{cavity}\approx8$ mm, which is in line with earlier measurements \citep{Zverkhovskyi2014}. Note that the estimated $h_{cavity}$ is much larger than the cavitator height, which is $h=2$ mm. 

Clearly, achieving a higher coverage would be beneficial, and could be achieved by measuring with a larger void fraction. Measurements at higher gas volume fractions are impossible due to vibrations of the system, which are caused by the uneven distribution of air.

\subsection{Torque and drag reduction}
\label{sec:torque}
Up to now, we focussed on flow visualizations and results which can be extracted from these. Now, we turn to torque measurements to quantify how the torque is affected by the air cavities. We study the influence of the Reynolds number, the void fraction $\alpha$, the number of cavitators and the effect of outer cylinder rotation on the global torque and drag reduction. In all measurements, we quasi-statically increase the rotation rates of the cylinders, and constantly measure the rotation rate, the torque, and the temperature in the flow. From the temperature we calculate the instantaneous viscosity and density.
The drag reduction is defined as $\text{DR} = 1 - G(\alpha) / G_{0}$. Here, we use as $G_{0}=G(\alpha=0)$ the case {\it without} air, but {\it with} cavitators. For the calculation of $G$ we used the density and viscosity of water, and we did not corrected for any changed effective flow properties caused by the air.

\begin{figure}[htp]
\centering
\includegraphics{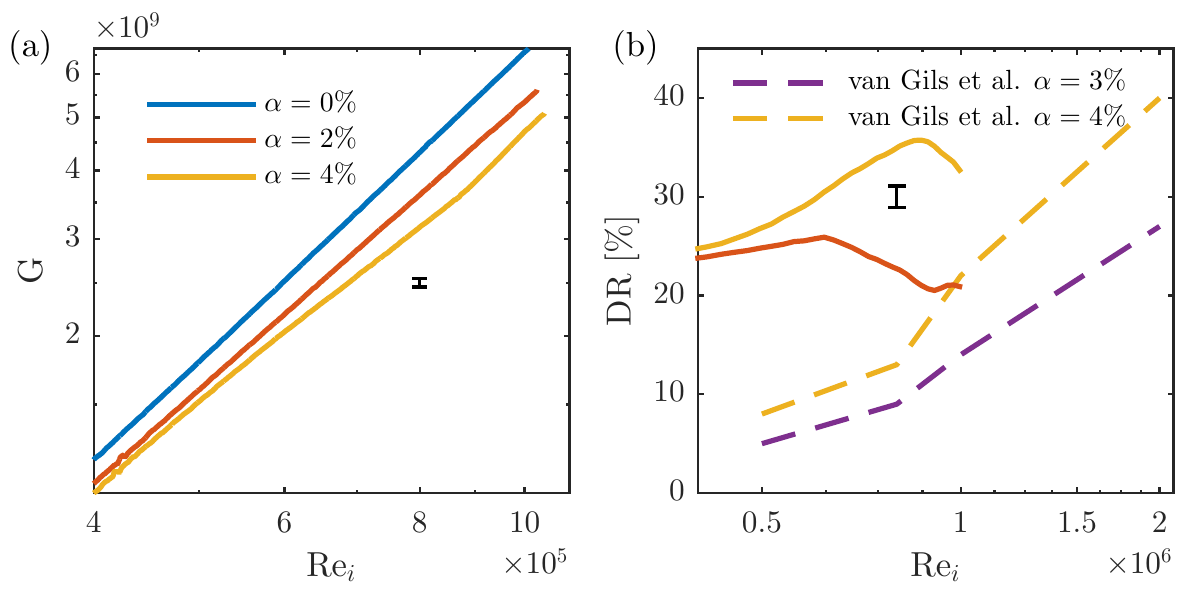}
\caption{Global dimensionless torque and drag reduction percentage for 0\%, 2\%, and 4\% of air. Here we mounted 3 cavitators, and we kept the outer cylinder stationary. {\bf (a)} Dimensionless torque $G$ as a function of the inner Reynolds number $\Re_i$.  {\bf (b)} Drag reduction percentages as a function of inner Reynolds number $\Re_i$. As comparison, we also show the bubbly DR results from \citet{gil13}. \red{A typical error bar is shown in both graphs.}}
\label{Chap_Six_fig:DR}
\includegraphics{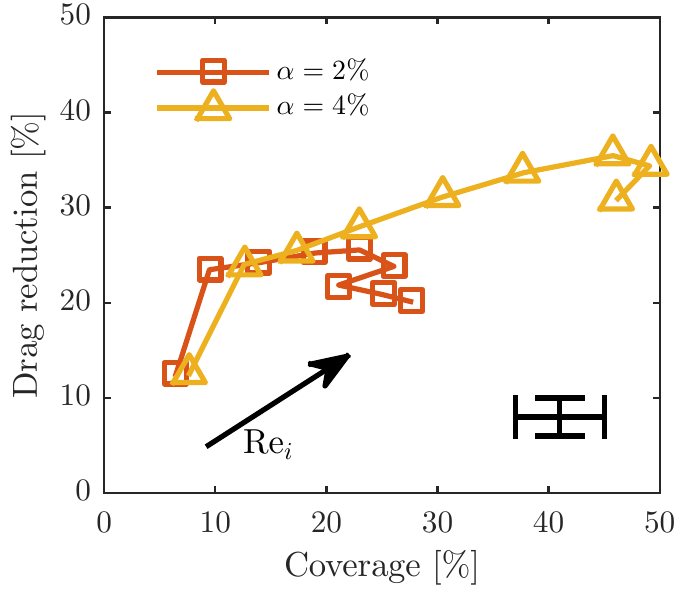}
\caption{\red{The drag reduction, as shown in fig.\ \ref{Chap_Six_fig:DR} as a function of air cavity coverage (fig.\ref{Chap_Six_fig:coverage}). Note that both the DR and the coverage depend on $\Re_i$. A typical error bar is shown for both the drag reduction as the coverage percentage.  }}
\label{Chap_Six_fig:DR_coverage}

\includegraphics{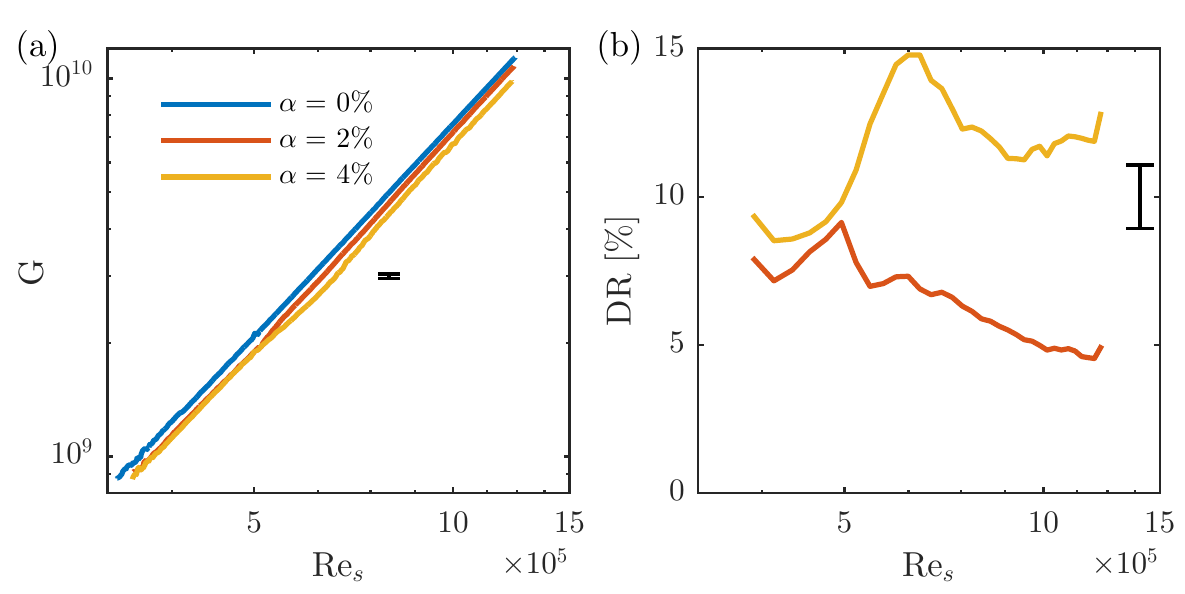}
\caption{Dimensionless torque and DR for the case of counter-rotating cylinders with 3 cavitators as a function of shear Reynolds number $\Re_s$. The rotation ratio equals $a=0.2$. 
{\bf (a)} Dimensionless torque $G$ as a function of shear Reynolds number $\Re_s$.  {\bf (b)} Drag reduction percentages as a function of shear Reynolds number $\Re_s$. The DR is significantly smaller than for the case of only inner cylinder rotation (fig.\ \ref{Chap_Six_fig:DR}). \red{A typical error bar is shown in both graphs.} }
\label{Chap_Six_fig:DR_CR}
\end{figure} 

We show results in fig.\ \ref{Chap_Six_fig:DR}. Here, we mount 3 cavitators to the IC, and while keeping $\alpha$ constant, we quasi-statically increase $\omega_i$, and thus $\Re_i$. The outer cylinder is kept stationary. We clearly see that the dimensionless torque $G$ decreases with the presence of air cavities. When we compare our air cavity results with earlier measurements with bubbly DR, we clearly observe that air cavities decrease the drag more effectively for the same Reynolds number. 
Nevertheless, we see that the air cavity DR saturates from $\Re_i=8\times10^5$ and onwards, whereas the bubbly DR increases with increasing Reynolds number. This can be explained as follows. When applying air cavities, the DR largely depends on the coverage, which is shown in fig.\ \ref{Chap_Six_fig:coverage}. We see in fig.\ \ref{Chap_Six_fig:coverage} that the air coverage saturates, which is reflected in the observed saturating DR. In fact, the shapes of fig.\ \ref{Chap_Six_fig:coverage} and \ref{Chap_Six_fig:DR}b are similar. \red{The relation between these two quantities is better revealed when plotting the DR as a function of air cavity coverage, see fig.\ \ref{Chap_Six_fig:DR_coverage}.}.
In bubbly DR, the bubbles do not necessarily attach to the cylinder. For bubbly DR, the relevant parameter is the Weber number, which is a measure for bubble deformability \citep{gil13}. The Weber number increases with increasing Reynolds number, hence the increasing DR. In the present study, the mechanism of DR is different, and as the coverage is limited by the void fraction, the DR consequently is limited too.

In the T$^3$C setup, we have the possibility to rotate the outer cylinder. It is known that for single-phase TC flow, the counterrotating cylinders enhance secondary flows, i.e.\ turbulent Taylor vortices \citep{gil12,ost14pd,gro16}. These Taylor vortices enhance the momentum transport from inner to outer cylinder, thus increasing the global torque \citep{gil11}.  By measuring in the counter-rotating regime we can study the influence of strong secondary flows on the cavity and the drag reduction.  We are not aware of any prior measurements of bubbles in turbulent TC flow with counterrotating cylinders. In these experiments we fix the rotation ratio between outer and inner cylinder to a rotation rate $a = -f_o/f_i=0.2$. 
The results are shown in fig.\ \ref{Chap_Six_fig:DR_CR}. Also in the counter-rotating regime the drag is decreased by air cavities. However, the DR is smaller than for pure inner cylinder rotation. This can be explained as follows.
In fig.\ \ref{Chap_Six_fig:visu2} we see that many bubbles are entrapped in the turbulent Taylor vortices, as the strong radial flow drags air away from the inner cylinder. Therefore, strong secondary flows decrease the stability of the air cavity, thus suppressing drag reduction. 

\begin{figure}[htp]
\centering]

\includegraphics[scale=1]{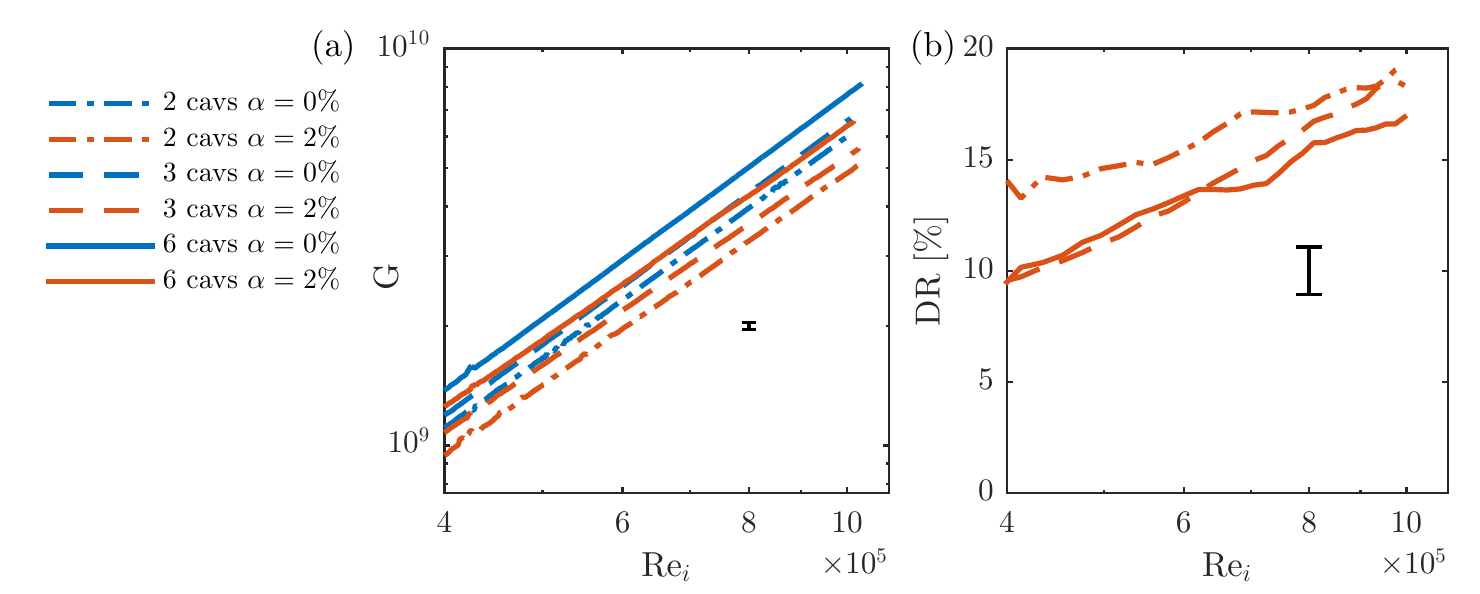}
\caption{{\bf (a)} Dimensionless torque with 2, 3, or 6 cavitators as a function of $\Re_i$ for stationary outer cylinder. {\bf (b)} The drag reduction for $\alpha=2\%$ for the case with 2, 3 or 6 cavitators. The DR percentages are similar for a constant gas volume fraction $\alpha$, although the global torque is increased by the cavitators, which induce an additional pressure drag \citep{zhu18}. \red{A typical error bar is shown in both graphs.} }
\label{Chap_Six_fig:DR_nrribs}

\includegraphics{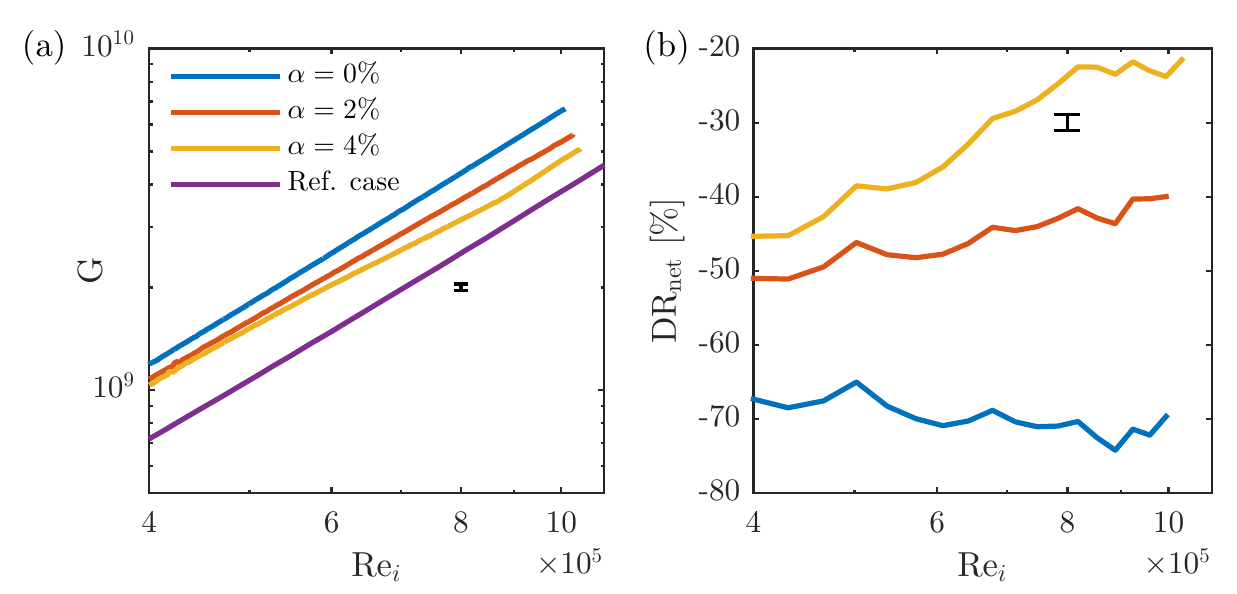}
\caption{Dimensionless torque and DR as a function of $\Re_i$. The reference case is without cavitators and with $\alpha=0$. The other cases are measured with 3 cavitators. The outer cylinder is stationary. {\bf (a)} The dimensionless torque as a function of the inner Reynolds number $\Re_i$. {\bf (b)} The net drag reduction as a function of the inner Reynolds number $\Re_i$. The net DR as compared to the reference case is negative, i.e.\ instead of drag reduction we observe a drag increase. \red{A typical error bar is shown in both graphs.}}
\label{Chap_Six_fig:DR_net}
\end{figure} 

We now vary the number of cavitators. We installed 2, 3, or 6 cavitators, and while keeping the outer cylinder stationary we measured the torque for void fractions of $\alpha=0\%$ and $\alpha=2\%$. The DR here is calculated comparing the case with $\alpha=2\%$ to the single-phase water case for the same number of cavitators. The results, which are shown in fig.\ \ref{Chap_Six_fig:DR_nrribs}, show that the DR is very similar. However, the absolute torque values  clearly differ, and increase with the number of riblets. E.g.\ here we see that the torque with 6 cavitators and $2\%$ void fraction is larger than the torque with 2 cavitators case without air. Since we clearly observe the additional drag caused by the cavitators, it is crucial to study the effect of the pressure drag at the cavitators in more detail.

As discussed in the introduction, one has to find the optimum between the DR caused by the cavities and the drag increase caused by the cavitators due to their pressure drag. In figures (\ref{Chap_Six_fig:DR}-\ref{Chap_Six_fig:DR_CR}) the presented DR percentages are relative to the case {\it without} air, but {\it with} cavitators. The presented DR values can be seen as a `gross drag reduction'. However, it is known that in TC flow (as for any other flow) even small roughness heights increase the drag tremendously \citep{ber03,zhu18}. The cavitators have a height of 2~mm, corresponding to 2.5\% of the gap width and to $\mathcal{O}( 10^2)$ wall units, so we are in the fully rough regime. In this paragraph, we study the effect of  these cavitators on the drag, by comparing our results with a reference case without cavitators. We define a `net DR' as: 
\begin{equation}
\text{DR}_\text{net} = 1 - G(\alpha) / G_{ref},
\end{equation}
in which $G_{ref}$ is a reference case {\it without} air and {\it without} cavitators, i.e.\ with smooth cylinders. The results are shown in fig.\ \ref{Chap_Six_fig:DR_net}. Clearly, the lowest dimensionless torque is obtained with the reference case. So,  when applying air cavities, the net drag is {\it increased}, as is also clear from fig.\ \ref{Chap_Six_fig:DR_net}b. We observe here that the drag increase caused by the cavitators is larger than the drag reduction caused by the air layer. We, however, note that measuring at larger void fractions and Reynolds numbers might cause $DR_{net}$ to be positive. Knowing the difference between net and gross drag reduction is crucial when applying air cavities. Full-scale ship experiments are extremely costly, and a reference test without cavitators might not be performed at all. Here, we show that the negative effects of the cavitators can be larger than the beneficial effects of the air cavity.

\begin{figure}[htp]
\centering]
\includegraphics[scale=1.2]{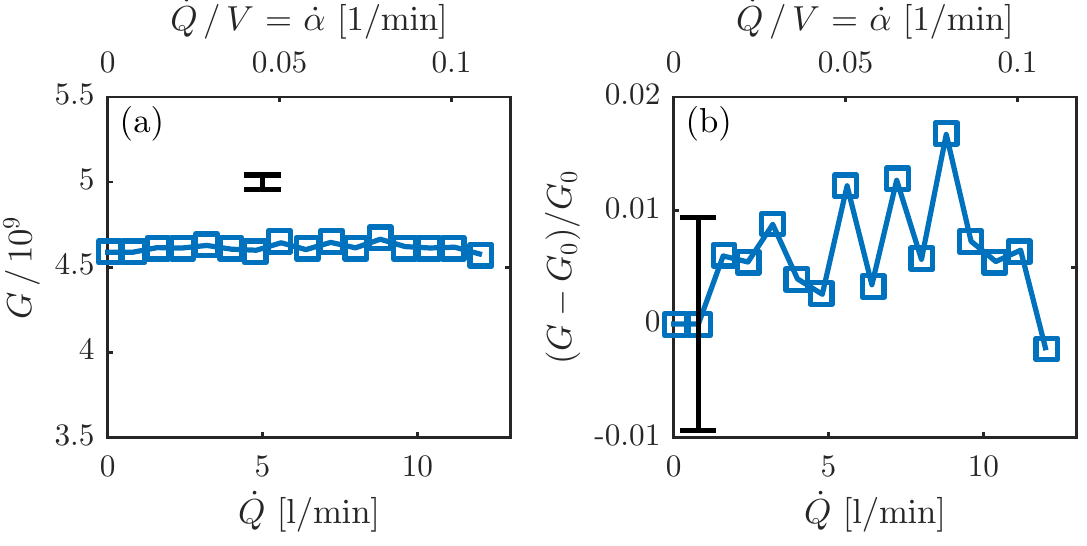}
\caption{{\bf (a)} The dimensionless torque $G$ as a function of gas flow rate $\dot{Q}$. We observe that $G$ does not depend on $\dot{Q}$. Here, the number of cavitators is 3, and we measure at $\Re_i=1\times 10^6$ with a void fraction of $\alpha=2\%$. The OC is kept stationary. On the top x-axis, we non-dimensialized $\dot{Q}$ with the volume of the system $V$, which equals $\dot{\alpha}$, i.e.\ the void fraction injected per minute. Injecting more air than shown here is not possible as it leads to an increase in void fraction as water is pushed out of the system, and thus to an unfair comparison. \red{In plot {\bf (b)}, we show the same data but made dimensionless: $(G-G_0)/G_0$, in which $G_0 = G(Q=0)$. In this way, the relative change of $G$ is shown. A typical error bar is shown in both graphs.}}
\label{Chap_Six_fig:DR_airinjection}
\end{figure} 

In water tunnel measurements the parameter governing the amount of air in the flow is the air injection rate $\dot{Q}$, which partially governs the cavity stability \citep{Zverkhovskyi2014}. $\dot{Q}$ should be sufficiently high to maintain the cavity, whereas a further increase in $\dot{Q}$ does not further increase the cavity length, but causes more air to be discharged in the closure region. 
In a closed TC system, air escaping from a certain cavitator can be re-entrained by other cavitators downstream. Similarly, in flat plate experiments it was shown that air which is discharged from a cavity can develop a new cavity if an additional cavitator is placed downstream \citep{Zverkhovskyi2014}. In open systems, such as ships or flat plates, the working fluid (with $\alpha=0\%$) is continuously refreshed such that $\dot{Q}$ is important, while the TC flow system is closed, meaning that $\alpha$ is the relevant parameter.
In all measurements presented above, we did not inject air locally, but instead chose to fill the cylinder only partially (see fig.\ \ref{Chap_Six_fig:setup}). Air is then entrained in the water by turbulent mixing. \red{It is not known to which extent active {\it local} gas injection influences the global torque and the air cavity length, given a certain $\alpha$.} 
The fact that pipe flow measurements showed that air can be reentrained at a cavitator placed further downstream indicates that rather than the air injection rate, the amount of available air is crucial \citep{Zverkhovskyi2014}. However, in flat plates it is not possible to disentangle air injection and local void fraction.
Here we study this by injection of a certain gas flow rate $\dot{Q}$, locally, directly at the cavitator, as indicated in fig.\ \ref{Chap_Six_fig:cav}, while keeping the void fraction constant at $\alpha=2\%$ and the Reynolds number constant at $\Re_i=1\times 10^6$. We simultaneously measure the torque. \red{We note that we inject significant amounts of air as compared to the amount of air which already is in the system (2.2 liter for $\alpha=2\%$), namely up to 5 times as much air per minute than the amount of air already present.} 

Surprisingly, we observe that active air injection does {\it not} influence the torque \red{in the studied range of air injection rates}, as shown in fig.\ \ref{Chap_Six_fig:DR_airinjection}. Apparently, the turbulent mixing in the flow is so strong that a steady state is reached almost immediately. Therefore, any excess of air is transported towards the top of the setup immediately, where it can leave the system. \red{This is somewhat similar to what was found in \citet{Zverkhovskyi2014}. In here, it was observed that increasing the gas flow in a flat plate setup does not increase the length of the cavity, and only leads to an increase in air discharge in the closure region.}

\section{Discussion and conclusions}
\label{sec:conclude}

A relevant question one could ask is: how can these Taylor-Couette results be compared with channel flow measurements \citep{murai2014,Foeth2008, cec10}? Several parameters which are common in the channel flow- and naval architecture communities are not used in TC flow and vice versa, see table \ref{tab:1} for a comparison of these parameters. As discussed above, as the air is not continuously swept away, like in an open system, it is $\alpha$, rather than 
$\dot{Q}$, that is the governing parameter in a closed flow system. The mean-field forcing of the flow is a second source of ambiguity. In TC flow, the mean-field forcing is not limited to the gravitational forces, as the centrifugal forces play a large role. The centrifugal forces $a_{centr} = \omega_i^2 r_i$ increase with Reynolds number until eventually,  in the limit of $\Re_i \rightarrow \infty$,  $a_{centr} \gg g$. As the centrifugal forces are directed in the radial direction, the water depth is to be taken in radial direction as well and thus equals the gap width $d$. Furthermore, in TC flow a second Froude number can be defined. The commonly used depth-based Froude number is  defined as $\Fr = \frac{u_{\infty}}{\sqrt{g D}}$, in which $u_{\infty}$ is a free stream velocity and $D$ is the water depth. Using the centrifugal acceleration we now define a ``centrifugal Froude number'' as $\Fr_{centr} = \frac{u_i}{\sqrt{a_{centr}d}} = \sqrt{r_i/d}= \sqrt{\eta / (1-\eta)}$, which, surprisingly, does not depend on the driving of the setup but only on the geometrical parameter $\eta$. In the currently used setup, the centrifugal Froude number equals $\Fr_{centr} = 1.59$.

Earlier studies showed that the length of the cavity equals half of the gravity wavelength, which is described by the dispersion relation $u_{\infty} = \sqrt{\frac{g \lambda}{2 \pi} \tanh\frac{2 \pi D}{\lambda}}$, in which $\lambda$ is the wavelength of the surface gravity waves \citep{butuzov1967,matveev2003}. Here, as $\Fr_{centr}>1$, the flow is supercritical, and the gravity wavelength becomes infinite \citep{Zverkhovskyi2014}. Thus, if enough air would be available \red{and for $a_{centr} \gg g$}, our streamwise cavity length would become unbounded and the entire cylinder is expected to be covered in air. \red{We note that in our study, both the buoyancy forces and centrifugal forces play a role, and therefore the flow is not yet supercritical, and consequently, the cylinder is not yet fully covered by cavities}.
\red{In channel flow, the cavitation number $\sigma$ is one of the basic parameters, and it is straightforward to measure. It is defined as $\sigma = (p - p_c )/ (\frac12 \rho u^2_{\infty})$, in which $p$ is the free stream pressure and $p_c$ is the pressure in the cavity. In TC flow, due to hydrostatic pressure, we define a height dependent cavity number as $\sigma_{TC} = (p(z) - p_c ) / (\frac12 \rho u^2_i)$.}

 A final difference between TC flow and channel flow is the way the systems are driven. Channel flow is pressure-driven, and consequently the momentum is transported from bulk to BLs. In TC flow, momentum is transported from the inner cylinder BL to the outer cylinder BL. The regimes of counter- and  co-rotation, caused by rotation of the outer cylinder are exclusive to the TC geometry.

\begin{table}[htp]
\begin{center}
\begin{tabular}{lll}
Quantity & Channel flow & TC flow \\
\hline
Amount of gas& $\dot{Q}$ [l/s] & $\alpha$ [\%] \\
Gravity & $\Fr$ &   For high $\Re_i$: $\Fr_{centr}$.  For low $\Re_i$: $\Fr$ \\
Water depth & D & For high $\Re_i$:  $r_o-r_i$. For low $\Re_i$: $L$ \\
\red{Cavitation number} & \red{$\sigma$} & \red{$\sigma_{TC}$}  \\
Driving & $\Re$ & $\Re_i$ and $\Re_o$  \\
\end{tabular}
\end{center}
\caption{Comparison between air cavity parameters for channel flow and TC flow.}
\label{tab:1}
\end{table}

Then, knowing these differences, what can be learned from these experiments, and how can these interpreted and applied by the naval industry? Our conclusion is that although one-to-one comparisons are difficult, the underlying physics remains the same. Therefore, our findings are of interest to anyone working on this topic.

To conclude, in this article we convincingly showed that air cavities can be (re)entrained in a Taylor-Couette flow setup. We show that air cavities result in gross DR percentages which are larger than the DR percentages for conventional bubble drag reduction. However, for all cases we see a net drag increase, caused by pressure drag at the cavitators. Therefore, when applying air cavities it is crucial to focus on the balance between drag reduction by the cavities and drag increase by the cavitators, closure region and any skegs. In addition, for maritime applications one should also take into consideration the energy costs to continuously inject air to judge whether or not a net gain can be achieved. 

We observed that the streamwise cavity length is significantly influenced by buoyancy effects. Therefore, we expect that air cavities on any non-flat bottomed hull behave similarly, and applying them is difficult. The global coverage is correlated to the Reynolds number and void fraction. To conclude, we showed that local air injection is not necessary, as long as sufficient amounts of air are available. This confirms that air which is discharged can be captured by any cavitator placed downstream on the hull.

In this exploratory study we restricted ourselves to one cavitator shape. Future work includes a study on the shape and size of the cavitators, preferably measuring at higher Reynolds numbers or at larger void fractions. The flow can be further quantified by local velocity measurements, which, although these are clearly difficult in multiphase flows, should be possible as the air is not dispersed homogeneously throughout the flow domain.

\begin{acknowledgments}
We thank Tom van Terwisga (MARIN, TU Delft) for the continuous and stimulating collaboration on drag reduction in the marine context over the years. We also thank Dennis van Gils, Gert-Wim Bruggert, and Martin Bos for their outstanding technical support. The work was financially supported by NWO-TTW (project 13265). \red{Huisman acknowledges support from MCEC. Sun and Bakhuis acknowledge financial support from VIDI grant No. 13477, and the Natural Science Foundation of China under grant no. 11672156. Bullee acknowledges NWO-TTW (project 14504).} The authors declare no conflicts of interests.
\end{acknowledgments}


\begin{thebibliography}{44}
\providecommand{\natexlab}[1]{#1}

\bibitem[{Amromin et~al.(2011)Amromin, Karafiath, and Metcalf}]{Amromin2011}
E.~Amromin, G.~Karafiath, and B.~Metcalf, Ship drag reduction by air bottom
  ventilated cavitation in calm water and waves, {J. Ship Research}
  \textbf{55}(3), 196--207 (2011).

\bibitem[{van~den Berg et~al.(2003)van~den Berg, Doering, Lohse, and
  Lathrop}]{ber03}
T.~H. van~den Berg, C.~R. Doering, D.~Lohse, and D.~P. Lathrop, Smooth and
  rough boundaries in turbulent {{Taylor-Couette}} flow, Phys. Rev. E
  \textbf{68}, 036307 (2003).

\bibitem[{van~den Berg et~al.(2007)van~den Berg, van Gils, Lathrop, and
  Lohse}]{ber07}
T.~H. van~den Berg, D.~P.~M. van Gils, D.~P. Lathrop, and D.~Lohse, Bubbly
  turbulent drag reduction is a boundary layer effect, Phys. Rev. Lett.
  \textbf{98}, 084501 (2007).

\bibitem[{van~den Berg et~al.(2005)van~den Berg, Luther, Lathrop, and
  Lohse}]{ber05}
T.~H. van~den Berg, S.~Luther, D.~P. Lathrop, and D.~Lohse, {Drag reduction in
  bubbly Taylor-Couette turbulence}, Phys. Rev. Lett. \textbf{94}, 044501
  (2005).

\bibitem[{Butuzov(1967)}]{butuzov1967}
A.~A. Butuzov, Artificial cavitation flow behind a slender wedge on the lower
  surface of a horizontal wall, Fluid Dyn. \textbf{2}(2), 56--58 (1967).

\bibitem[{Ceccio(2010)}]{cec10}
S.~L. Ceccio, Friction drag reduction of external flows with bubble and gas
  injection, Annu. Rev. Fluid Mech. \textbf{42}, 183--203 (2010).

\bibitem[{Chouippe et~al.(2014)Chouippe, Climent, Legendre, and
  Gabillet}]{cho14}
A.~Chouippe, E.~Climent, D.~Legendre, and C.~Gabillet, Numerical simulation of
  bubble dispersion in turbulent {{Taylor-Couette}} flow, Phys. Fluids
  \textbf{26}(4), 043304 (2014).

\bibitem[{Climent et~al.(2007)Climent, M., and Magnaudet}]{climent2007}
E.~Climent, S.~M., and J.~Magnaudet, Preferential accumulation of bubbles in
  {Couette-Taylor} flow patterns, {Phys. Fluids} \textbf{19}, 083301 (2007).

\bibitem[{Djeridi et~al.(2004)Djeridi, Gabillet, and Billard}]{Djeridi2004}
H.~Djeridi, C.~Gabillet, and J.~Y. Billard, {Two-phase Couette-Taylor flow:
  Arrangement and affects on the flow structures}, {Phys. Fluids} \textbf{16},
  128 (2004).

\bibitem[{Elbing et~al.(2013)Elbing, M\"{a}kiharju, Wiggins, Perlin, Dowling,
  and Ceccio}]{elbing2013}
B.~R. Elbing, S.~M\"{a}kiharju, A.~Wiggins, M.~Perlin, D.~R. Dowling, and S.~L.
  Ceccio, {On the scaling of air layer drag reduction}, J. Fluid Mech.
  \textbf{717}, 484--513 (2013).

\bibitem[{Elbing et~al.(2008)Elbing, Winkel, Lay, Ceccio, Dowling, and
  Perlin}]{elbing2008}
B.~R. Elbing, E.~S. Winkel, K.~A. Lay, S.~L. Ceccio, D.~R. Dowling, and
  M.~Perlin, {Bubble-induced skin-friction drag reduction and the abrupt
  transition to air-layer drag reduction}, J. Fluid Mech. \textbf{612},
  201--236 (2008).

\bibitem[{Fardin et~al.(2014)Fardin, Perge, and Taberlet}]{far14}
M.~A. Fardin, C.~Perge, and N.~Taberlet, {``}{{The}} hydrogen atom of fluid
  dynamics{"} - {{Introduction}} to the {{Taylor-Couette}} flow for {{Soft
  Matter}} scientists, Soft Matter \textbf{10}, 3523 (2014).

\bibitem[{Foeth(2008)}]{Foeth2008}
E.~J. Foeth, Decreasing frictional resistance by air lubrication, in 20th
  International HISWA Symposium on Yacht Design and Yacht Construction (HISWA,
  2008).

\bibitem[{van Gils et~al.(2011)van Gils, Huisman, Bruggert, Sun, and
  Lohse}]{gil11}
D.~P.~M. van Gils, S.~G. Huisman, G.~W. Bruggert, C.~Sun, and D.~Lohse, Torque
  scaling in turbulent {{Taylor-Couette}} flow with co- and counter-rotating
  cylinders, Phys. Rev. Lett. \textbf{106}, 024502 (2011).

\bibitem[{van Gils et~al.(2012)van Gils, Huisman, Grossmann, Sun, and
  Lohse}]{gil12}
D.~P.~M. van Gils, S.~G. Huisman, S.~Grossmann, C.~Sun, and D.~Lohse, Optimal
  {{Taylor-Couette}} turbulence, J. Fluid Mech. \textbf{706}, 118--149 (2012).

\bibitem[{van Gils et~al.(2013)van Gils, {{Narezo Guzman}}, Sun, and
  Lohse}]{gil13}
D.~P.~M. van Gils, D.~{{Narezo Guzman}}, C.~Sun, and D.~Lohse, The importance
  of bubble deformability for strong drag reduction in bubbly turbulent
  {{Taylor-Couette}} flow, J. Fluid Mech. \textbf{722}, 317--347 (2013).

\bibitem[{Grossmann et~al.(2016)Grossmann, Lohse, and Sun}]{gro16}
S.~Grossmann, D.~Lohse, and C.~Sun, High {{Reynolds}} number {{Taylor-Couette}}
  turbulence, Ann. Rev. Fluid Mech. \textbf{48}, 53 (2016).

\bibitem[{Hoang et~al.(2009)Hoang, Toda, and Sanada}]{Hoang2009}
C.~Hoang, Y.~Toda, and Y.~Sanada, Full scale experiment for frictional
  resistance reduction using air lubrication method, in Proc. of the 19th
  International Offshore and Polar Engineering Conference, 812--817 (ISOPE,
  2009).

\bibitem[{Kumagai et~al.(2015)Kumagai, Takahashi, and Murai}]{kumagai2015}
I.~Kumagai, Y.~Takahashi, and Y.~Murai, {Power-saving device for air bubble
  generation using a hydrofoil to reduce ship drag: Theory, experiments, and
  application to ships}, Ocean Eng. \textbf{95}, 183--194 (2015).

\bibitem[{Larsson and Raven(2010)}]{larsson2010}
L.~Larsson and H.~C. Raven, The principles of naval architecture series: ship
  resistance and flow (The Society of Naval Architects and Marine Engineers,
  New York, 2010).

\bibitem[{Latorre(1997)}]{Latorre1997}
R.~Latorre, Ship hull drag reduction using bottom air injection, Ocean Eng.
  \textbf{24}(2), 161--175 (1997).

\bibitem[{Lay et~al.(2010)Lay, Yakushiji, M{\"a}kiharju, Perlin, and
  Ceccio}]{Lay2010}
K.~A. Lay, R.~Yakushiji, S.~M{\"a}kiharju, M.~Perlin, and S.~L. Ceccio, Partial
  cavity drag reduction at high {Reynolds} numbers, {J. Ship Research}
  \textbf{52}(2), 109--119 (2010).

\bibitem[{Lu et~al.(2005)Lu, Fernandez, and Tryggvason}]{lu05}
J.~C. Lu, A.~Fernandez, and G.~Tryggvason, Drag reduction in a turbulent
  channel due to bubble injection, Phys. Fluids \textbf{17}, 095102 (2005).

\bibitem[{M{\"a}kiharju et~al.(2013)M{\"a}kiharju, Elbing, Wiggins, Schinasi,
  Vanden-Broeck, Perlin, Dowling, and Ceccio}]{Makiharju2013}
S.~M{\"a}kiharju, B.~R. Elbing, A.~Wiggins, S.~Schinasi, J.-M. Vanden-Broeck,
  M.~Perlin, D.~R. Dowling, and S.~L. Ceccio, On the scaling of air entrainment
  from a ventilated partial cavity, {J. Fluid Mech.} \textbf{732}(47-76)
  (2013).

\bibitem[{M{\"a}kiharju et~al.(2012)M{\"a}kiharju, Perlin, and
  Ceccio}]{Makiharju2012}
S.~M{\"a}kiharju, M.~Perlin, and S.~L. Ceccio, On the energy economics of air
  lubrication drag reduction, {Int. J. of Naval Architecture in Oceanic Eng.}
  \textbf{4}, 412--422 (2012).

\bibitem[{Matveev(2003)}]{matveev2003}
K.~I. Matveev, On the limiting parameters of artificial cavitation, Ocean Eng.
  \textbf{30}(9), 1179--1190 (2003).

\bibitem[{Matveev(2005)}]{matveev2005}
K.~I. Matveev, Applications of artificial cavitation for reducing ship drag,
  Ocean Eng. Int. \textbf{9}(1), 35--41 (2005).

\bibitem[{Mizokami et~al.(2010)Mizokami, Kawakita, Kodan, Takano, Higasa, and
  Shigenaga}]{Mizokami2010}
S.~Mizokami, C.~Kawakita, Y.~Kodan, S.~Takano, S.~Higasa, and R.~Shigenaga,
  Experimental study of air lubrication method and verification of effects on
  actual hull by means of sea trial, Mitsubishi Heavy Industries Technical
  Review \textbf{47}(93), 41--47 (2010).

\bibitem[{Murai(2014)}]{murai2014}
Y.~Murai, {Frictional drag reduction by bubble injection}, Exp. Fluids
  \textbf{55}(7), 1773 (2014).

\bibitem[{Murai et~al.(2005)Murai, Oiwa, and Takeda}]{mur05}
Y.~Murai, H.~Oiwa, and Y.~Takeda, Bubble behavior in a vertical
  {{Taylor-Couette}} flow, J. Phys. (Conf. Series) \textbf{14}, 143--156
  (2005).

\bibitem[{Murai et~al.(2008)Murai, Oiwa, and Takeda}]{mur08}
Y.~Murai, H.~Oiwa, and Y.~Takeda, Frictional drag reduction in bubbly
  {Couette--Taylor} flow, {Phys. Fluids} \textbf{20}(3), 034101 (2008).

\bibitem[{Ndongo~Fokoua et~al.(2015)Ndongo~Fokoua, Gabillet, Aubert, and
  Colin}]{fokoua2015}
G.~Ndongo~Fokoua, C.~Gabillet, A.~Aubert, and C.~Colin, Effect of bubble's
  arrangement on the viscous torque in bubbly {Taylor-Couette} flow, {Phys.
  Fluids} \textbf{27}, 034105 (2015).

\bibitem[{Ostilla-M\'onico et~al.(2014)Ostilla-M\'onico, van~der Poel,
  Verzicco, Grossmann, and Lohse}]{ost14pd}
R.~Ostilla-M\'onico, E.~P. van~der Poel, R.~Verzicco, S.~Grossmann, and
  D.~Lohse, Exploring the phase diagram of fully turbulent {{Taylor-Couette}}
  flow, J. Fluid Mech. \textbf{761}, 1--26 (2014).

\bibitem[{Rosenberg et~al.(2016)Rosenberg, van Buren, Matthew, and
  Smits}]{ros16}
B.~J. Rosenberg, T.~van Buren, K.~F. Matthew, and A.~J. Smits, Turbulent drag
  reduction over air- and liquid- impregnated surfaces, {Phys. Fluids}
  \textbf{28}, 015103 (2016).

\bibitem[{Rotte et~al.(2016)Rotte, Zverkhovskyi, Kerkvliet, and van
  Terwisga}]{Rotte2016}
G.~M. Rotte, O.~Zverkhovskyi, M.~Kerkvliet, and T.~J.~C. van Terwisga, On the
  physical mechanisms for the numerical modelling of flows around air
  lubricated ships, in {Proc. 12th Int. Conf. on Hydrodynamics} (ICHD, 2016).

\bibitem[{van Ruymbeke et~al.(2017)van Ruymbeke, Murai, Tasaka, Oishi,
  Gabillet, and Colin}]{Ruymbeke2017}
B.~van Ruymbeke, Y.~Murai, Y.~Tasaka, Y.~Oishi, C.~Gabillet, and C.~Colin,
  Quantitative visualization of swirl and cloud bubbles in {Taylor--Couette}
  flow, {J. Visualization} \textbf{20}, 349 (2017).

\bibitem[{Sanders et~al.(2006)Sanders, Winkel, Dowling, Perlin, and
  Ceccio}]{san06}
W.~C. Sanders, E.~S. Winkel, D.~R. Dowling, M.~Perlin, and S.~L. Ceccio, Bubble
  friction drag reduction in a high-reynolds-number flat-plate turbulent
  boundary layer, J. Fluid Mech. \textbf{552}, 353--380 (2006).

\bibitem[{Saranadhi et~al.(2016)Saranadhi, Chen, Kleingartner, Srinivasan,
  Cohen, and McKinley}]{sar16}
D.~Saranadhi, D.~Chen, J.~A. Kleingartner, S.~Srinivasan, R.~B. Cohen, and
  G.~H. McKinley, Sustained drag reduction in a turbulent flow using a low
  temperature {Leidenfrost} surface, {Science Advances} \textbf{2}(10),
  E1600686 (2016).

\bibitem[{Spandan et~al.(2016)Spandan, Ostilla-M\'{o}nico, Verzicco, and
  Lohse}]{Spandan2016}
V.~Spandan, R.~Ostilla-M\'{o}nico, R.~Verzicco, and D.~Lohse, {Drag reduction
  in numerical two-phase Taylor--Couette turbulence using an Euler--Lagrange
  approach}, {J. Fluid. Mech.} \textbf{798}, 411--435 (2016).

\bibitem[{Srinivasan et~al.(2015)Srinivasan, Kleingartner, Gilbert, Cohen,
  Milne, and McKinley}]{sri15}
S.~Srinivasan, J.~A. Kleingartner, J.~B. Gilbert, R.~B. Cohen, A.~J.~B. Milne,
  and G.~H. McKinley, Sustainable drag reduction in turbulent {Taylor-Couette}
  flows by depositing sprayable superhydrophobic surfaces, {Phys. Rev. Lett.}
  \textbf{114}, 014501 (2015).

\bibitem[{Sugiyama et~al.(2008)Sugiyama, Calzavarini, and Lohse}]{sug08b}
K.~Sugiyama, E.~Calzavarini, and D.~Lohse, Microbubble drag reduction in
  {{Taylor-Couette}} flow in the wavy vortex regime, J. Fluid Mech.
  \textbf{608}, 21--41 (2008).

\bibitem[{Verschoof et~al.(2016)Verschoof, van~der Veen, Sun, and
  Lohse}]{ver16}
R.~A. Verschoof, R.~C.~A. van~der Veen, C.~Sun, and D.~Lohse, {Bubble drag
  reduction requires large bubbles}, {Phys. Rev. Lett.} \textbf{117}, 104502
  (2016).

\bibitem[{Zhu et~al.(2018)Zhu, Verschoof, Bakhuis, Huisman, Verzicco, Sun, and
  Lohse}]{zhu18}
X.~Zhu, R.~A. Verschoof, D.~Bakhuis, S.~G. Huisman, R.~Verzicco, C.~Sun, and
  D.~Lohse, Wall roughness induces asymptotic ultimate turbulence., Nat. Phys.
  \textbf{14}, 417--423 (2018).

\bibitem[{Zverkhovskyi(2014)}]{Zverkhovskyi2014}
O.~Zverkhovskyi, Ship drag reduction by air cavities, Ph.D. thesis, Delft
  University of Technology, Delft, NL (2014).

\end{thebibliography}

\end{document}